\documentclass[useAMS]{mn2e}

\usepackage{graphicx}
\title[Vega's rotational velocity]
{Determination of Vega's rotational velocity \\
based on the Fourier analysis of spectral line profiles
}

\author[Y. Takeda]
{Yoichi Takeda\thanks{E-mail:
ytakeda@js2.so-net.ne.jp}\footnotemark[0] 
\\
11-2 Enomachi, Naka-ku, Hiroshima-shi 730-0851, Japan\\
}
\begin{document}

\date{Accepted 2021 May 11. Received 2021 May 11; in original form 2021 April 11}

%\pagerange{\pageref{firstpage}--\pageref{lastpage}} \pubyear{2002}

\maketitle

\label{firstpage}

\begin{abstract}
While it is known that the sharp-line star Vega ($v_{\rm e}\sin i \sim 20$~km~s$^{-1}$) 
is actually a rapid rotator seen nearly pole-on with low $i$ $(< 10^{\circ})$, no consensus 
has yet been accomplished regarding its intrinsic rotational velocity ($v_{\rm e}$), 
for which rather different values have been reported so far. 
Methodologically, detailed analysis of spectral line profiles is useful for this purpose, 
since they reflect more or less the $v_{\rm e}$-dependent gravitational darkening effect. 
However, direct comparison of observed and theoretically simulated line profiles 
is not necessarily effective in practice, where the solution is sensitively affected 
by various conditions and the scope for combining many lines is lacking. 
In this study, determination of Vega's $v_{\rm e}$ was attempted based on 
an alternative approach making use of the first zero ($q_{1}$) of 
the Fourier transform of each line profile, which depends upon $K$ 
(temperature sensitivity parameter differing from line to line) and $v_{\rm e}$.
It turned out that $v_{\rm e}$ and $v_{\rm e}\sin i$ could be separately established by 
comparing the observed $q_{1}^{\rm obs}$ and calculated $q_{1}^{\rm cal}$ 
values for a number of lines of different $K$. Actually, independent analysis applied to 
two line sets (49 Fe~{\sc i} lines and 41 Fe~{\sc ii} lines) yielded results reasonably 
consistent with each other. The final parameters of Vega's rotation were concluded as  
$v_{\rm e}\sin i = 21.6 (\pm 0.3)$~km~s$^{-1}$,
$v_{\rm e} = 195 (\pm 15)$~km~s$^{-1}$, and $i = 6.4 (\pm 0.5)^{\circ}$.
\end{abstract}


%Sect. 1
\section{Introduction}

The spectrum of Vega (= $\alpha$~Lyr = HR~7001 = HD~172167 = HIP~91262; spectral type 
A0~V) shows a sharp-line nature indicating a small projected rotational velocity 
($v_{\rm e}\sin i \sim 20$~km~s$^{-1}$; where $v_{\rm e}$ is the equatorial rotation 
velocity and $i$ is the angle of rotational axis relative to the line of sight), 
which is rather unusual among A-type main-sequence stars (many of them showing 
$v_{\rm e}\sin i$ typically around $\sim$~100--300~km~s$^{-1}$).
It is nowadays known, however, that this star is actually a rapid rotator with large 
$v_{\rm e}$ like other A stars and the apparent smallness of $v_{\rm e}\sin i$ is 
simply ascribed to low $i$ (i.e., this star happens to be seen nearly pole-on). 

Its intrinsic rotational velocity can be observationally determined by detecting 
the gravity darkening effect, because it becomes more exaggerated as $v_{\rm e}$ increases. 
The mainstream approach used for this purpose is to analyse the shape of 
spectral lines, because lines of a specific group (e.g., weak Fe~{\sc i} lines) show 
a characteristic feature (i.e., flat-bottomed profile), which is caused by the lowered 
temperature near to the gravity-darkened limb (see, e.g., Fig.~1 in Takeda, Kawanomoto \& Ohishi 2008a). 
Alternatively, in order to establish $v_{\rm e}$, the extent of gravity darkening 
can be estimated from the brightness distribution on the stellar disk by direct 
high-resolution interferometric observations.

%Table 1
\setcounter{table}{0}
\begin{table*}
\begin{minipage}{180mm}
\small
%\scriptsize
\caption{Previous determinations of Vega's rotation and related parameters.}
\begin{center}
\begin{tabular}{cccccccl}\hline
\hline
Authors & $v_{\rm e}\sin i$ & $v_{\rm e}$ & $i$ & $R_{\rm p}$ & $R_{\rm e}$ & $P$ & Remark \\ 
        &  (km~s$^{-1}$)  &  (km~s$^{-1}$) & (deg) & (${\rm R}_{\odot}$) & (${\rm R}_{\odot}$) & (d) & \\
\hline
Gulliver et al. (1994) & 21.8 & 245 & 5.1  & $\cdots$ & $\cdots$ & $\cdots$ & Line profile \\
Hill et al. (2004) & 21.9 & 160 & 7.9  & $\cdots$ & $\cdots$ & $\cdots$ & Line profile \\
Aufdenberg et al. (2006) & 21.9 & 270 & 4.7  & 2.26  & 2.78  & $\cdots$ & Interferometry \\ 
Peterson et al. (2006) & 21.5 & 274 & 4.5  & 2.31  & 2.87  & $\cdots$ & Interferometry \\ 
Takeda et al. (2008b) & $^{*}$22 & 175 & 7.2  & 2.52  & 2.76  & $\cdots$ & Line profile \\ 
Yoon et al. (2010) & 20.5 & 236 & 5.0  & 2.36  & 2.82  & $\cdots$ & Line profile \\ 
Hill et al. (2010) & 20.8 & 211 & 5.7  & 2.40  & 2.75  & $\cdots$ & Line profile \\
Monnier et al. (2012) & 21.3 & 197 & 6.2  & 2.42  & 2.73  & $\cdots$ & Interferometry (their Model~3) \\
\hline
Petit et al. (2010) & $\cdots$ & $^{\dagger}$184 & $\cdots$ & $\cdots$ & $\cdots$ & 0.732 & Magnetic modulation  \\
Alina et al. (2012) & $\cdots$ & $^{\dagger}$198 & $\cdots$ & $\cdots$ & $\cdots$ & 0.678 & Magnetic modulation  \\ 
Butkovskaya (2014) & $\cdots$ & $^{\dagger}$216 & $\cdots$ & $\cdots$ & $\cdots$ & 0.623 & Magnetic modulation \\
B\"{o}hm et al. (2015) & $\cdots$ & $^{\dagger}$198 & $\cdots$ & $\cdots$ & $\cdots$ & 0.678 & Magnetic modulation  \\ 
\hline
\end{tabular}
\end{center}
In columns 2--7 are given the values of projected rotational velocity, equatorial rotational velocity,
inclination angle of rotational axis, polar radius, equatorial radius, and rotation period, respectively.\\
$^{*}$Assumed value.\\
$^{\dagger}$Derived from $P$ by assuming $R_{\rm e} = 2.8 {\rm R}_{\odot}$
\end{minipage}
\end{table*}

Beginning from 1990s and especially in the period around 2010, quite a few 
determinations of Vega's $v_{\rm e}$ based on these two methods have been 
tried by various investigators as summarised in Table~1.
However, the resulting literature values of $v_{\rm e}$ considerably differ from 
each other as seen from this table. Although the large discrepancy amounting
to $\ga 100$~km~s$^{-1}$ (from $\sim 160$ to $\sim 270$~km~s$^{-1}$) seen 
in early 2000s has been mitigated up to the present, they are still diversified
between $\sim 170$~km~s$^{-1}$ and $\sim 230$~km~s$^{-1}$ (which are the published 
results since 2008). 

Meanwhile, the discovery of magnetic field in Vega by spectropolarimetry 
(Ligni\`{e}res et al. 2009) provided a new means to measure $v_{\rm e}$, 
because such a Zeeman signature would show cyclic variation due to rotation.
That is, the rotational period ($P$) may be directly evaluated by applying a period 
analysis to time-series data of spectropolarimetric observations, from which $v_{\rm e}$
is derived as $v_{\rm e} = 2\pi R_{\rm e}/P$ by using an appropriately assigned 
$R_{\rm e}$ (equatorial radius).
Following this policy, Vega's rotation period was determined within several years 
after 2010, as shown in Table~1. Although this method is expected to establish $P$
precisely, these published data are not necessarily in good agreement
but somewhat discrepant by $\sim\pm 10$\% (i.e., $\sim \pm 20$~km~s$^{-1}$ around
$v_{\rm e} \sim 200$~km~s$^{-1}$). Therefore, even such an independent technique 
(which is essentially different from the other in the sense that any modelling of 
gravity-darkened star is not required) has not yet significantly improved 
the situation regarding the ambiguity in $v_{\rm e}$.

Accordingly, it is desirable to redetermine $v_{\rm e}$ of Vega with higher reliability
than before, in order to clarify which of the recent results (between ``low-scale'' value  
of $\sim$~170--180~km~s$^{-1}$ and ``high-scale'' value of  $\sim$~220--230~km~s$^{-1}$)
is more justifiable.

Here, it may be worthwhile to mention the weakpoint of line profile analysis, 
which was once employed by the author's group (Takeda, Kawanomoto \& Ohishi 2008b; 
hereinafter referred to as Paper~I) to evaluate Vega's $v_{\rm e}$. 
According to our experience, to derive $v_{\rm e}$ by searching for the best fit (minimising 
$\chi^{2}$) between the observed and modelled line profiles for a selected line feature 
(e.g., well-behaved weak Fe~{\sc i} line showing a flat-bottomed profile) is not so hard.
However, there is no way to estimate how much uncertainty is involved in such a
specific solution. Actually, since $\chi^{2}$ residual is a rather broad function of $v_{\rm e}$ 
and quite vulnerable to a slight imperfection (e.g., improper placement of continuum level,
existence of weak line blending, irregular noise in observed data, etc.), because
extremely subtle difference of profile shape is concerned (typically on the order 
of $\sim 10^{-3}$ in unit of the continuum; cf. Figs. 4 and 5 in Paper~I),
an erroneous $v_{\rm e}$ solution is easily brought about (or even no solution
is found). Therefore, it was decided in Paper~I to analyse the profiles of 
a large number of lines (87 lines of neutral species and 109 lines of once-ionised 
species) with a hope of hitting as many correct solutions as possible.  
Nevertheless, from a critical point of view, the result obtained 
in Paper~I was not very satisfactory for the following reasons: (i) The final 
solution ($v_{\rm e} = 175$~km~s$^{-1}$) was simply selected from 9 models (where 
$v_{\rm e}$ was varied from 100 to 300~km~s$^{-1}$ with an increment of 25~km~s$^{-1}$) 
as the one corresponding to the highest frequency of $\chi^{2}$ minimum for the case 
of neutral lines; so an ambiguity of $\sim$~20~km~s$^{-1}$ 
due to the coarseness of model grid is inevitable from the start. (ii) While lines of 
neutral species yielded a Gaussian-like frequency histogram centred around 
175~km~s$^{-}$ (cf. Fig.~6a in Paper~I), those of ionised species (many of them 
have ``non-flat-bottom'' profiles) show a near-flat distribution (cf. Fig.~6b in 
Paper~I); this means that the latter set of ionised lines were almost useless 
because they made no contribution to the determination of $v_{\rm e}$. 

Consequently, the conventional line-profile matching in the wavelength domain applied
in Paper~I was not necessarily suitable for such a very delicate problem. In order to make 
a further step towards improving the precision, a more efficient approach has to be invoked,
in which many lines of different properties can be effectively combined to increase
the reliability of $v_{\rm e}$ solution while providing a reasonable procedure for 
error estimation. 

Recently, in an attempt to estimate the intrinsic rotational velocity of Sirius~A, 
Takeda (2020; hereinafter referred to as Paper~II) made use of the first zero frequency
($q_{1}$) in the Fourier transform of the line profile. It then revealed that
this quantity can be used for measuring the gravity darkening effect because it
sensitively responds to a slight variation of the line profile; actually, $q_{1}$ was 
found to be vary almost monotonically with $v_{\rm e}$ (inducing a gravity darkening).
While how $q_{1}$ reflects a change of $v_{\rm e}$ naturally differs from line to line 
depending on its property, it was found to be the sensitivity of line strength
($W$) to temperature ($T$), which is represented by the parameter $K (\equiv \log W/\log T)$,
that essentially controls the $v_{\rm e}$-dependence of $q_{1}$. 
Therefore, since information of $v_{\rm e}$ may be extracted from the comparison of the 
observed $q_{1}^{\rm obs}$ with a corresponding set of $q_{1}^{\rm cal}(K,v_{\rm e}$ 
calculated for this line on the models of different $v_{\rm e}$, the best solution 
of $v_{\rm e}$ (along with its probable error) can be established by combining 
many lines of different $K$.
This technique turned out successful, and in Paper~II was concluded that Sirius~A
is an intrinsically slow rotator ($16 \le v_{\rm e} \la$~30--40~km~s$^{-1}$). 

Motivated by this achievement, the author decided to apply this method to analysing 
the spectral line profiles of Vega, in order to revisit the task of determining its 
$v_{\rm e}$ as done in Paper~I, hoping that a result of higher accuracy would be 
obtained, so that the diversified literature values may be verified. 
The purpose of this article is to report the outcome of this reinvestigation.

%Sect. 2
\section{Observational data}

\subsection{Selection of lines and their profiles}

Regarding the basic observational material of Vega, the high-dispersion spectra of 
high signal-to-noise ratio (S/N~$\sim 2000$) and high spectral resolving 
power ($R \sim 100000$) were used as in Paper~I, which were obtained at Okayama 
Astrophysical Observatory by using the HIDES spectrograph attached to the 188~cm 
reflector and published by Takeda, Kawanomoto \& Ohishi (2007).

The selection of lines to be used for the analysis was done by following almost the 
same procedure as adopted in Paper~II (cf. Sect.~2.2 therein), where it was decided 
to employ only lines of neutral and ionised Fe in order to maintain consistency with 
Paper~II. As a result, a total of 90 lines (49 Fe~{\sc i} and 41 Fe~{\sc ii} lines)    
were eventually sorted out,\footnote{Since the selection criterion adopted in this
study differs from that of Paper~I, the resulting line set is somewhat different.
More precisely, out of 60/52 Fe~{\sc i}/Fe~{\sc ii} lines analysed in Paper~I,
17/16 were discarded, while 6/5 were newly included.} which are listed in Table~2.  
The observed profiles of these lines are displayed in Fig.~1, and their original
data are available in ``obsprofs.dat'' of the supplementary material.
 
The equivalent widths ($W^{\rm obs}$) of these 90 lines were measured by 
the Gaussian fitting method, which are in the range of 1~m\AA~$\la W^{\rm obs} \la 40$~m\AA.
As the ``standard''  plane-parallel model atmosphere for Vega, Kurucz's (1993) ATLAS9 
model with $T_{\rm eff}$ = 9630~K, $\log g = 3.94$, $v_{\rm t} = 2$~km~s$^{-1}$ 
(microturbulence), and [X/H]~=~$-0.5$ (metallicity) was adopted in this study as in Paper~I,
which well reproduces the spectral energy distribution. By using this model 
along with the atomic data taken from Kurucz \& Bell's (1995) compilation, the abundance 
($A^{\rm std}$; called as ``standard abundance'') was derived from $W^{\rm obs}$ 
for each line.

In the same manner as in Paper~II (cf. Sect.~4.1 therein), the $T$-sensitivity 
parameter $K (\equiv {\rm d}\log W/{\rm d}\log T)$ was then evaluated as 
\begin{equation}
K \equiv \frac{(W^{+100} - W^{-100})/W^{\rm obs}}{(+100-(-100))/9630},
\end{equation}
where $W^{+100}$ and $W^{-100}$ are the equivalent widths computed from $A^{\rm std}$  
by two model atmospheres with only $T_{\rm eff}$ being perturbed by $+100$~K 
($T_{\rm eff} = 9730$~K) and  $-100$~K ($T_{\rm eff} = 9530$~K), respectively
(while other parameters are kept the same as the standard values).
The ranges of the resulting $K$ values are (roughly) $-20 \la K \la -10$ and   
$-5 \la K \la +5$ for Fe~{\sc i} and Fe~{\sc ii} lines, respectively.

\subsection{Zero frequencies of Fourier transforms}

Then, the Fourier transform $d(\sigma)$ of the line depth profile 
$D_{\lambda} (\equiv 1 - F_{\lambda}/F_{\rm cont})$ was calculated for each line
as done in Paper~II (cf. Sect.~2.3 therein), and
the 1st and 2nd zero frequencies ($\sigma_{1}$ and $\sigma_{2}$; in unit of wavelength) 
were measured from the cuspy features of $|d(\sigma)|$, which were further converted
to wavelength-independent quantities ($q_{1}$ and $q_{2}$; in unit of velocity$^{-1}$)
for convenience by the relation $q \equiv \sigma /(c \lambda)$ ($c$: velocity of light).
The resulting $q_{1}$ and $q_{2}$ are plotted against the line parameters in Fig.~2,
from which the following arguments can be made. 
\begin{itemize}
\item
These zero frequencies show an appreciable line-dependent scatter; especially,
those of a fraction of Fe~{\sc ii} lines are remarkably higher in comparison 
with the theoretical values expected from the classical rotational broadening 
function (cf. Fig.~2a)
\item
This implies that the conventional Fourier analysis of spectral line profiles, 
which assumes that the observed profile is expressed by a convolution of the 
rotational broadening function and thus the zero frequency of the rotational 
broadening function (dependent upon $v_{\rm e}\sin i$) should be simply inherited 
in the observed transform equally for any line, is no more feasible for precise 
$v_{\rm e} \sin i$ determination in this case.
\item
The cause for this scatter in $q_{1}$ as well as $q_{2}$ is that they tend to 
systematically increase with $K$ as shown in Fig.~2b. This is because the 
line profile characteristics is determined by this $T$-sensitivity parameter.
That is, a line of small/negative $K$ (e.g., weak Fe~{\sc i} line of low excitation) 
shows a boxy {\bf U}-shape, while that of large/positive $K$ (e.g., 
weak Fe~{\sc ii} line of high excitation) has a sharp {\bf V}-shape.  
Such a difference in the line profile (even if very subtle) is reflected by 
the position of zero frequency, which is actually verified by theoretical 
calculations based on the gravity-darkened rotating star model (cf. Sect.~3.3).
\item
These $q_{1}$ and $q{2}$ also show some systematic trends with respect to 
$\chi_{\rm low}$ (Fig.~2c) and $W^{\rm obs}$ (Fig.~2d); but they can be 
reasonably explained by the dependence of $K$ upon $\chi_{\rm low}$ 
and $W^{\rm obs}$, as discussed in Appendix~A2 of Paper~II.
Accordingly, it is the difference in $K$ that causes the line-by-line 
different characteristics in the profile (and the zero positions).
\end{itemize}

The atomic line data and the values of $W^{\rm obs}$, $K$, $q_{1}$, and $q_{2}$ for 
90 lines are presented in Table~2. Besides, more complete data (including $A^{\rm std}$
and the main lobe height as well as the 1st sidelobe height) are summarised in 
``obsparms.dat'' of the supplementary material.

%Fig. 1
\setcounter{figure}{0}
\begin{figure*}
\begin{minipage}{130mm}
\includegraphics[width=13.0cm]{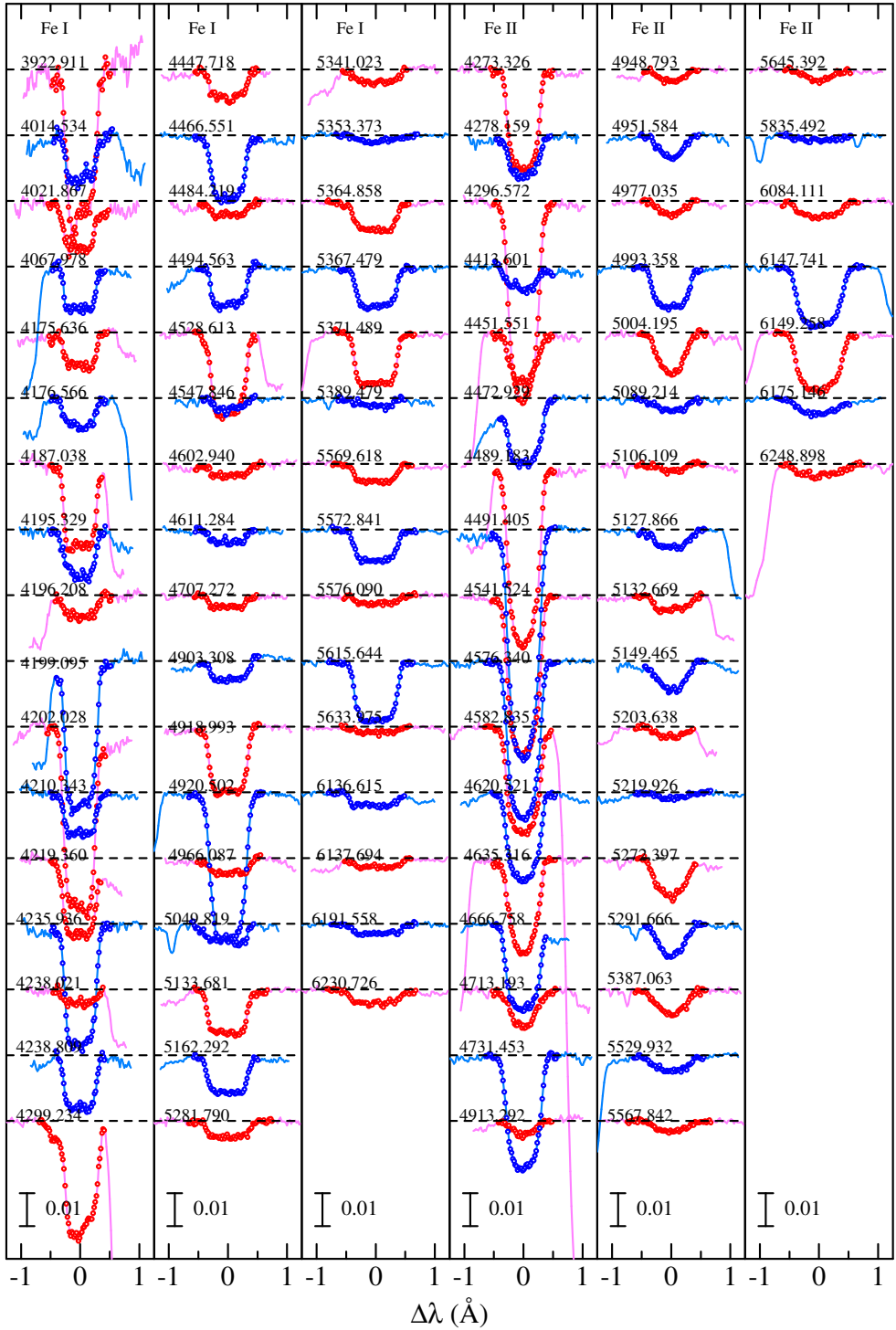}
\caption{
Observed spectra of finally selected 49 Fe~{\sc i} lines (1st through 3rd panels) 
and 41 Fe~{\sc ii} lines (4th through 6th panels), which are arranged in the 
increasing order of wavelength within each group of species as in Table~2. 
The actual spectral data (normalised flux plotted against the wavelength 
displacement relative to the line centre) are shown by lines, while the selected wavelength 
portions [$\lambda_{1}$, $\lambda_{2}$] used for calculating the Fourier transforms 
are depicted by symbols. Each spectrum (its continuum level is indicated by the horizontal
dashed line) is shifted by 0.02 (2\% of the continuum level) 
relative to the adjacent one.
}
\label{fig1}
\end{minipage}
\end{figure*}

%Fig. 2
\setcounter{figure}{1}
\begin{figure*}
\begin{minipage}{120mm}
\includegraphics[width=12.0cm]{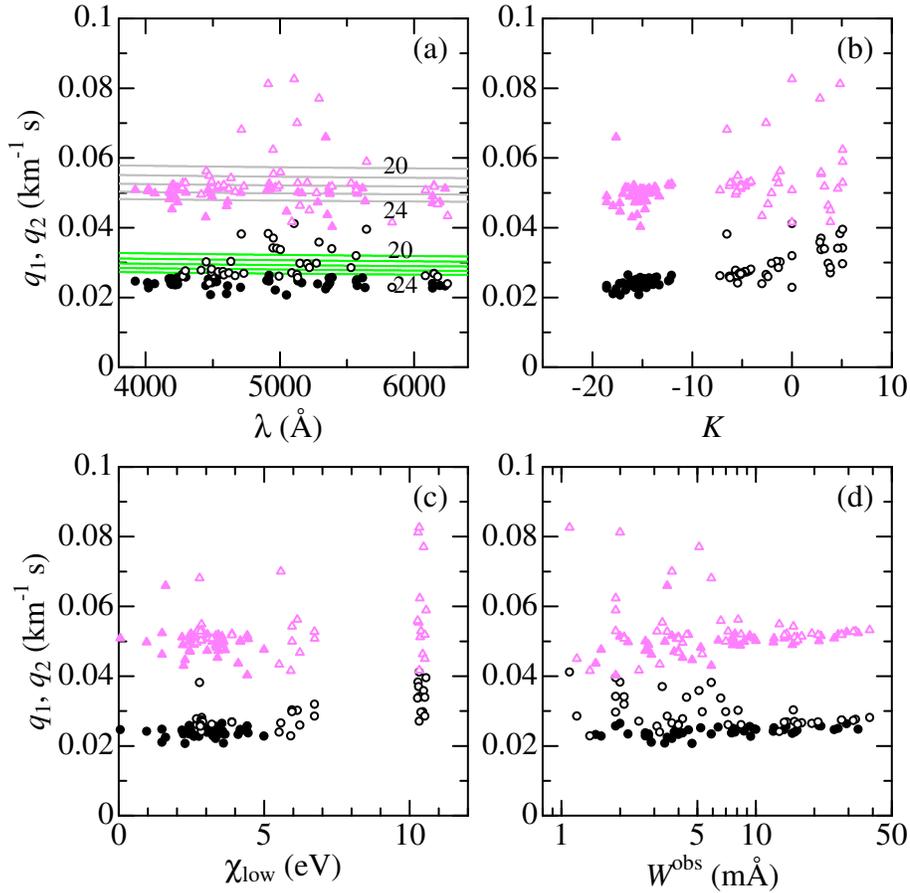}
\caption{
The 1st-zero frequencies ($q_{1}$: circles) and 2nd-zero frequencies 
($q_{2}$: triangles) of Fourier transforms, which were calculated 
from the observed profiles of 90 Fe lines, are plotted against
(a) $\lambda$ (wavelength), (b) $K$ ($T$-sensitivity parameter), (c) $\chi_{\rm low}$
(lower excitation potential), and $W^{\rm obs}$ (observed equivalent width).
The filled and open symbols correspond to Fe~{\sc i} and Fe~{\sc ii} lines, respectively. 
In panel~(a), the classical $q_{1}(\lambda)$ and $q_{2}(\lambda)$ values
derived from the conventional rotational broadening function corresponding to 
$v_{\rm e}\sin i$ = 20, 21, 22, 23, and 24~km~s$^{-1}$ (which were derived 
from Eqs.~3, 4, and 6 in Paper~II) are depicted by solid lines. 
}
\label{fig2}
\end{minipage}
\end{figure*}

%Table 2
\setcounter{table}{1}
\begin{table}
%\begin{minipage}{90mm}
%\small
\scriptsize
\caption{Atomic data and observed quantities of adopted spectral lines.}
\begin{center}
\begin{tabular}{crcrccc}\hline
\hline
$\lambda$ & $\chi_{\rm low}$ & $\log gf$ & $W^{\rm obs}$ & $K$ & $q_{1}^{\rm obs}$ & $q_{2}^{\rm obs}$ \\
 (\AA) & (eV) & (dex) & (m\AA) &  & (km$^{-1}$s) &  (km$^{-1}$s) \\
\hline
\multicolumn{7}{c}{(49 Fe~{\sc i} lines)}\\
 3922.911 & 0.052 & $-$1.651 & 25.3 & $-$14.89 & 0.02465 & 0.05085 \\
 4014.534 & 3.573 & $-$0.200 &  7.6 & $-$14.48 & 0.02425 & 0.05050 \\
 4021.867 & 2.759 & $-$0.660 &  9.4 & $-$15.21 & 0.02276 & 0.05120 \\
 4067.978 & 3.211 & $-$0.430 &  7.8 & $-$14.63 & 0.02393 & 0.04988 \\
 4175.636 & 2.845 & $-$0.670 &  6.4 & $-$14.82 & 0.02540 & 0.05027 \\
 4176.566 & 3.368 & $-$0.620 &  5.2 & $-$15.59 & 0.02517 & 0.04823 \\
 4187.038 & 2.449 & $-$0.548 & 15.9 & $-$14.36 & 0.02398 & 0.04980 \\
 4195.329 & 3.332 & $-$0.412 &  8.3 & $-$14.90 & 0.02538 & 0.04882 \\
 4196.208 & 3.396 & $-$0.740 &  4.2 & $-$14.73 & 0.02371 & 0.04530 \\
 4199.095 & 3.047 & +0.250 & 26.5 & $-$12.26 & 0.02503 & 0.05200 \\
 4202.028 & 1.485 & $-$0.708 & 33.6 & $-$12.38 & 0.02477 & 0.05236 \\
 4210.343 & 2.482 & $-$0.870 &  7.8 & $-$15.85 & 0.02426 & 0.05217 \\
 4219.360 & 3.573 & +0.120 & 15.4 & $-$13.31 & 0.02362 & 0.04897 \\
 4235.936 & 2.425 & $-$0.341 & 21.6 & $-$13.69 & 0.02568 & 0.05155 \\
 4238.021 & 3.417 & $-$1.286 &  2.7 & $-$15.76 & 0.02373 & 0.04730 \\
 4238.809 & 3.396 & $-$0.280 &  9.8 & $-$14.59 & 0.02490 & 0.05084 \\
 4299.234 & 2.425 & $-$0.430 & 21.1 & $-$13.56 & 0.02584 & 0.05010 \\
 4447.718 & 2.223 & $-$1.342 &  5.9 & $-$16.05 & 0.02342 & 0.04301 \\
 4466.551 & 2.832 & $-$0.590 & 12.4 & $-$14.96 & 0.02520 & 0.05088 \\
 4484.219 & 3.602 & $-$0.720 &  3.4 & $-$15.35 & 0.02078 & 0.04992 \\
 4494.563 & 2.198 & $-$1.136 &  7.6 & $-$16.26 & 0.02423 & 0.04895 \\
 4528.613 & 2.176 & $-$0.822 & 16.3 & $-$14.93 & 0.02450 & 0.05115 \\
 4547.846 & 3.546 & $-$0.780 &  2.0 & $-$16.44 & 0.02646 & 0.05141 \\
 4602.940 & 1.485 & $-$1.950 &  2.9 & $-$17.95 & 0.02107 & 0.04621 \\
 4611.284 & 3.654 & $-$0.670 &  2.8 & $-$15.21 & 0.02335 & 0.04747 \\
 4707.272 & 3.241 & $-$1.080 &  2.7 & $-$15.76 & 0.02289 & 0.05004 \\
 4903.308 & 2.882 & $-$1.080 &  3.9 & $-$15.85 & 0.02487 & 0.05074 \\
 4918.993 & 2.865 & $-$0.370 & 14.0 & $-$14.24 & 0.02469 & 0.05088 \\
 4920.502 & 2.832 & +0.060 & 29.3 & $-$12.08 & 0.02636 & 0.05265 \\
 4966.087 & 3.332 & $-$0.890 &  3.7 & $-$16.69 & 0.02211 & 0.05162 \\
 5049.819 & 2.279 & $-$1.420 &  4.7 & $-$17.23 & 0.02071 & 0.04464 \\
 5133.681 & 4.178 & +0.140 & 10.0 & $-$13.76 & 0.02473 & 0.05017 \\
 5162.292 & 4.178 & +0.020 &  8.7 & $-$14.23 & 0.02422 & 0.04991 \\
 5281.790 & 3.038 & $-$1.020 &  4.0 & $-$15.46 & 0.02387 & 0.04728 \\
 5341.023 & 1.608 & $-$2.060 &  3.5 & $-$17.63 & 0.02246 & 0.06592 \\
 5353.373 & 4.103 & $-$0.840 &  1.5 & $-$15.53 & 0.02327 & 0.04369 \\
 5364.858 & 4.446 & +0.220 &  7.5 & $-$13.94 & 0.02370 & 0.05068 \\
 5367.479 & 4.415 & +0.350 &  9.3 & $-$13.76 & 0.02557 & 0.05168 \\
 5371.489 & 0.958 & $-$1.645 & 12.3 & $-$16.95 & 0.02419 & 0.04971 \\
 5389.479 & 4.415 & $-$0.410 &  1.9 & $-$15.21 & 0.02572 & 0.04028 \\
 5569.618 & 3.417 & $-$0.540 &  4.5 & $-$15.87 & 0.02460 & 0.05192 \\
 5572.841 & 3.396 & $-$0.310 &  7.6 & $-$15.01 & 0.02508 & 0.05084 \\
 5576.090 & 3.430 & $-$1.000 &  2.2 & $-$14.98 & 0.02342 & 0.04979 \\
 5615.644 & 3.332 & $-$0.140 & 14.3 & $-$14.33 & 0.02563 & 0.05133 \\
 5633.975 & 4.991 & $-$0.270 &  1.6 & $-$14.59 & 0.02282 & 0.04763 \\
 6136.615 & 2.453 & $-$1.400 &  3.8 & $-$17.28 & 0.02402 & 0.04708 \\
 6137.694 & 2.588 & $-$1.403 &  2.8 & $-$18.58 & 0.02271 & 0.04905 \\
 6191.558 & 2.433 & $-$1.600 &  2.8 & $-$18.58 & 0.02341 & 0.04736 \\
 6230.726 & 2.559 & $-$1.281 &  4.0 & $-$16.44 & 0.02328 & 0.05124 \\
\hline
\end{tabular}
\end{center}
%\end{minipage}
\end{table}

%Table 2
\setcounter{table}{1}
\begin{table}
%\begin{minipage}{90mm}
%\small
\scriptsize
\caption{(Continued.)}
\begin{center}
\begin{tabular}{crcrccc}\hline
\hline
 $\lambda$ & $\chi_{\rm low}$ & $\log gf$ & $W^{\rm obs}$ & $K$ & $q_{1}^{\rm obs}$ & $q_{2}^{\rm obs}$ \\
 (\AA) & (eV) & (dex) & (m\AA) &  & (km$^{-1}$s) &  (km$^{-1}$s) \\
\hline
\multicolumn{7}{c}{(41 Fe~{\sc ii} lines)}\\
 4273.326 & 2.704 & $-$3.258 & 17.2 &  $-$5.30 & 0.02662 & 0.05143 \\
 4278.159 & 2.692 & $-$3.816 &  7.2 &  $-$5.98 & 0.02611 & 0.05284 \\
 4296.572 & 2.704 & $-$3.010 & 32.6 &  $-$4.28 & 0.02764 & 0.05285 \\
 4413.601 & 2.676 & $-$3.870 &  4.2 &  $-$5.66 & 0.02780 & 0.04958 \\
 4451.551 & 6.138 & $-$1.844 &  8.1 &  $-$1.19 & 0.03020 & 0.05623 \\
 4472.929 & 2.844 & $-$3.430 & 13.2 &  $-$5.45 & 0.02412 & 0.05491 \\
 4489.183 & 2.828 & $-$2.970 & 31.7 &  $-$4.54 & 0.02730 & 0.05278 \\
 4491.405 & 2.855 & $-$2.700 & 38.5 &  $-$4.12 & 0.02813 & 0.05322 \\
 4541.524 & 2.855 & $-$3.050 & 27.7 &  $-$4.68 & 0.02737 & 0.05199 \\
 4576.340 & 2.844 & $-$3.040 & 27.6 &  $-$4.70 & 0.02744 & 0.05185 \\
 4582.835 & 2.844 & $-$3.100 & 19.4 &  $-$5.20 & 0.02645 & 0.05223 \\
 4620.521 & 2.828 & $-$3.280 & 15.8 &  $-$5.16 & 0.02702 & 0.05118 \\
 4635.316 & 5.956 & $-$1.650 & 15.6 &  $-$1.54 & 0.03032 & 0.05428 \\
 4666.758 & 2.828 & $-$3.330 & 16.4 &  $-$5.56 & 0.02623 & 0.05034 \\
 4713.193 & 2.778 & $-$4.932 &  5.9 &  $-$6.53 & 0.03818 & 0.06809 \\
 4731.453 & 2.891 & $-$3.360 & 20.9 &  $-$5.29 & 0.02692 & 0.05184 \\
 4913.292 &10.288 & +0.012 &  2.0 &  +4.82 & 0.03830 & 0.08123 \\
 4948.793 &10.347 & $-$0.008 &  1.9 &  +5.07 & 0.03419 & 0.06236 \\
 4951.584 &10.307 & +0.175 &  3.3 &  +2.92 & 0.03701 & 0.05541 \\
 4977.035 &10.360 & +0.041 &  2.1 &  +4.59 & 0.03409 & 0.05122 \\
 4993.358 & 2.807 & $-$3.650 &  8.4 &  $-$6.27 & 0.02610 & 0.05199 \\
 5004.195 &10.272 & +0.497 &  6.6 &  +2.92 & 0.03372 & 0.05589 \\
 5089.214 &10.329 & $-$0.035 &  2.5 &  +3.85 & 0.02708 & 0.04168 \\
 5106.109 &10.329 & $-$0.276 &  1.1 &   0.00 & 0.04118 & 0.08262 \\
 5127.866 & 5.570 & $-$2.535 &  3.7 &  $-$2.60 & 0.02657 & 0.07004 \\
 5132.669 & 2.807 & $-$4.180 &  3.1 &  $-$6.21 & 0.02566 & 0.05284 \\
 5149.465 &10.447 & +0.396 &  5.3 &  +3.63 & 0.02978 & 0.04636 \\
 5203.638 &10.391 & $-$0.046 &  1.9 &  +5.07 & 0.02968 & 0.05292 \\
 5219.926 &10.522 & $-$0.366 &  1.2 &  +3.85 & 0.02856 & 0.04502 \\
 5272.397 & 5.956 & $-$2.030 &  7.0 &  $-$2.05 & 0.02980 & 0.04995 \\
 5291.666 &10.480 & +0.575 &  5.1 &  +2.80 & 0.03584 & 0.07705 \\
 5387.063 &10.521 & +0.518 &  4.4 &  +3.25 & 0.03395 & 0.05187 \\
 5529.932 & 6.729 & $-$1.875 &  3.5 &  $-$1.36 & 0.02854 & 0.05276 \\
 5567.842 & 6.730 & $-$1.887 &  2.1 &   0.00 & 0.03195 & 0.05079 \\
 5645.392 &10.561 & +0.085 &  1.9 &  +5.07 & 0.03955 & 0.05893 \\
 5835.492 & 5.911 & $-$2.372 &  1.4 &   0.00 & 0.02289 & 0.04158 \\
 6084.111 & 3.199 & $-$3.808 &  4.0 &  $-$7.22 & 0.02622 & 0.05076 \\
 6147.741 & 3.889 & $-$2.721 & 14.4 &  $-$5.00 & 0.02691 & 0.05213 \\
 6149.258 & 3.889 & $-$2.724 & 14.0 &  $-$5.14 & 0.02678 & 0.05177 \\
 6175.146 & 6.222 & $-$1.983 &  4.0 &  $-$2.41 & 0.02596 & 0.04678 \\
 6248.898 & 5.511 & $-$2.696 &  3.2 &  $-$3.01 & 0.02396 & 0.04336 \\
\hline
\end{tabular}
\end{center}
In columns 1--7 are given the line wavelength, lower excitation potential, 
logarithm of oscillator strength times lower level's statistical weight, 
observed equivalent width, $T$-sensitivity parameter, observed 1st zero.
frequency, and observed 2nd zero frequency, respectively. The atomic data 
are taken from the compilation of Kurucz \& Bell (1995).
%\end{minipage}
\end{table}

%Sect. 3
\section{Modelling of line profiles}

\subsection{Adopted model parameters}

Regarding the simulation of theoretical line profiles of a gravity-darkened rotating 
star, this study follows the same assumptions and procedures (including the adopted set 
of parameters for Vega) as described in Paper~I, where the stellar mass ($M$), rotational 
velocity at the equator ($v_{\rm e}$), inclination angle of rotation axis ($i$),
polar radius ($R_{\rm p}$), and polar effective temperature ($T_{\rm eff,p}$) 
are the fundamental parameters to be specified.

The mass was fixed at $M = 2.3$~M$_{\odot}$, Ten $v_{\rm e}$ values were chosen as
22, 100, 125, 150, $\cdots$ 275, and 300~km~s$^{-1}$, (numbered as models 
0, 1, 2, 3, $\cdots$, 8, and 9), and the corresponding $i$ values were derived from the 
assumption of $v_{\rm e}\sin i = 22$~km~s$^{-1}$ (which is a reasonable value for Vega).
Based on the requirement of spectral energy distribution, $R_{\rm p}$ and $T_{\rm eff,p}$
can be expressed as 2nd-order polynomials in terms of $v_{\rm e}$ (cf. Eqs. 1 and 2 in Paper~I).
The model parameters for each of the 10 models are summarised in Table~3, 
which is the same as Table~1 in Paper~I. 
Note that model~0 ($v_{\rm e} = 22$~km~s$^{-1}$ and $i = 90^{\circ}$) is a special model
different from others, in the sense that it is a spherically symmetric rigid model 
where the gravity effect (darkening and distortion) is intentionally suppressed.
This model~0 is almost equivalent to the ``standard model'' mentioned in Sect.~2.1. 

%Table 3
\setcounter{table}{2}
\begin{table*}
\begin{minipage}{180mm}
\small
%\scriptsize
\caption{Parameters of adopted models for rotating Vega.}
\begin{center}
\begin{tabular}{cccccccccl}\hline
Model & $v_{\rm e}$ & $i$ & $R_{\rm p}$ & $R_{\rm e}$ & $T_{\rm eff,p}$ & $T_{\rm eff,e}$ & $\log g_{\rm p}$ & $\log g_{\rm e}$  & Remark \\
number & (km~s$^{-1}$) & (deg) & (${\rm R}_{\odot}$) & (${\rm R}_{\odot}$) & (K) & (K) & (cm~s$^{-2}$) & (cm~s$^{-2}$) & \\ 
\hline
0 &  22 & 90.0 & 2.700 & 2.700 &  9630 & 9630 & 3.937 & 3.937 & Gravity effect suppressed.\\
1 & 100 & 12.7 & 2.640 & 2.722 &  9698 & 9399 & 3.956 & 3.956 & \\
2 & 125 & 10.1 & 2.600 & 2.726 &  9750 & 9281 & 3.969 & 3.884 & \\
3 & 150 &  8.4 & 2.560 & 2.740 &  9806 & 9126 & 3.983 & 3.858 & \\
4 & 175 &  7.2 & 2.520 & 2.763 &  9867 & 8931 & 3.997 & 3.823 & Nominated model in Paper~I. \\ 
5 & 200 &  6.3 & 2.470 & 2.784 &  9932 & 8695 & 4.014 & 3.783 & Best model concluded in this study. \\
6 & 225 &  5.6 & 2.410 & 2.799 & 10000 & 8416 & 4.035 & 3.736 & \\
7 & 250 &  5.0 & 2.360 & 2.837 & 10074 & 8072 & 4.054 & 3.669 & \\
8 & 275 &  4.6 & 2.300 & 2.869 & 10151 & 7787 & 4.076 & 3.587 & \\
9 & 300 &  4.2 & 2.240 & 2.908 & 10233 & 7546 & 4.099 & 3.477 & \\
\hline
\end{tabular}
\end{center}
Given are the model number, equatorial rotation velocity, inclination angle, radius, 
effective temperature, and logarithmic surface gravity at the pole as well as the equator.  
These models are the same as adopted in Paper~I (cf. Table~1 therein).
Note that $v_{\rm e}\sin i$ is assumed to be 22~km~s$^{-1}$ in all these models.
\end{minipage}
\end{table*}

\subsection{Simulation of line profiles}

The emergent line flux profile was simulated with the program CALSPEC (cf. Sect.~4.1 
in Paper~I) by integrating the intensity profile at each point on the visible disk, 
which was generated by using the local model atmosphere corresponding to 
$T_{\rm eff}(\Theta)$, $g(\Theta)$, $v_{\rm t}$~=~2~km~s$^{-1}$,
and [X/H] = $-0.5$ (where $\Theta$ is the co-latitude).

Here, a point to notice is how to assign the elemental abundance ($A$). 
If $A^{\rm std}$ (standard abundance derived from the classical plane-parallel model) 
is simply used, the equivalent width of the calculated line profile ($W^{\rm cal}$) 
turns out generally stronger than $W^{\rm obs}$ because of the gravity darkening 
effect,\footnote{Although $A^{\rm std}$ was simply used in Paper~II for all models 
irrespective of $v_{\rm e}$,it did not cause any serious problem because 
$v_{\rm e}$-dependent gravity darkening effect was not so large as to cause 
a significant $W^{\rm cal}$ vs. $W^{\rm obs}$ discrepancy in the $v_{\rm e}$ 
range ($\le 150$~km~s$^{-1}$) inspected therein.}  
and this discrepancy progressively increases towards higher $v_{\rm e}$
(as can be recognised in Figs. 4 and 5 in Paper~I).
In Paper~I, this problem was circumvented by renormalising the calculated profile
(cf. Eq. 7 therein) so as to force $W^{\rm cal} = W^{\rm obs}$, although its validity
was not necessarily clear.

Fortunately, this equality does not have to be strictly realised in the present case
of Fourier analysis, because it is the ``characteristics'' of the line shape 
that is essential. Accordingly, the following procedure was adopted in this study.
\begin{itemize}
\item
First, the provisional equivalent width $W_{*}^{j}$ ($j = 0, 1, 2, \cdots, 9$) 
was calculated with CALSPEC for each model by using $A^{\rm std}$. 
\item
Then, the corresponding abundance $A_{*}^{j}$ was derived from $W_{*}^{j}$ 
with the help of Kurucz's (1993) WIDTH9 program by using the standard plane 
parallel model (cf. Sect. 2.1).
\item
The abundance difference defined as $\Delta A^{j} \equiv A^{\rm std} - A_{*}^{j}$
(which is mostly negative) is used as abundance correction to be applied to $A^{\rm std}$.
That is the abundance actually adopted in CALSPEC for calculating the profile 
corresponding to model~$j$ is $A^{\rm std} + \Delta A^{j}$. 
\end{itemize}
It should be remarked that this procedure is based on two assumptions that
(i) the classical curve of growth ($\log W$ vs. $A$ relation) for the plane-parallel model 
is applicable even for the gravity-darkened case, and (ii) the absolute change of 
$\log W$ in response to perturbation by $\pm \Delta A$ around $A^{\rm std}$ 
in this curve of growth is almost the same (i.e., locally linear).
Despite these rough approximations, the discrepancy between $W^{\rm cal}$ and 
$W^{\rm obs}$ seen for the case of simply using $A^{\rm std}$ is considerably 
reduced by application of this correction ($\Delta A$), as shown in Fig.~3.

%Fig. 3
\setcounter{figure}{2}
\begin{figure}
%\begin{minipage}{80mm}
\includegraphics[width=8.0cm]{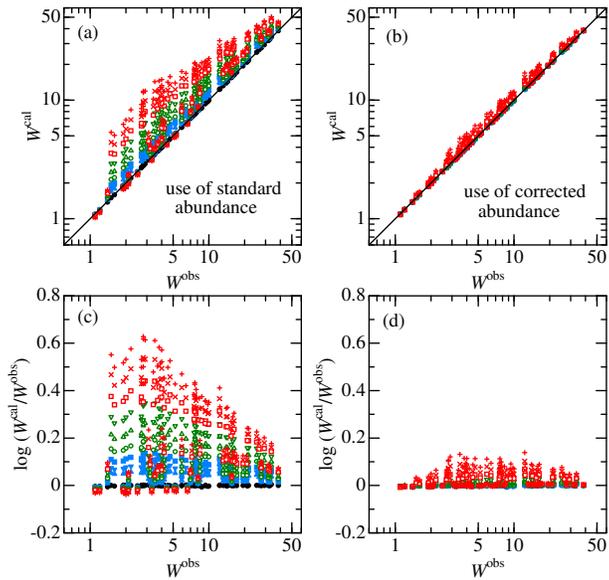}
\caption{
Graphical illustration describing how the abundance correction applied to the 
standard abundance (cf. Sect.~3.2) improves the discrepancy between the observed 
($W^{\rm obs}$) and calculated ($W^{\rm cal}$) equivalent widths for each line, 
where $W^{\rm cal}$ (upper panels) and $\log (W^{\rm cal}/W^{\rm obs})$ (lower panels) 
are plotted against $W^{\rm obs}$. The left-hand panels (a, c) correspond to
the case of using the standard (uncorrected) abundances, while the right-hand panels
(b, d) to the case of using the corrected abundances. The results for models 0, 1, 2, 3,
4, 5, 6, 7, 8, and 9 are denoted by filled circles, filled triangles, filled inverse triangles,
filled squares, open circles, open triangles, open inverse triangles, open squares,
St. Andrew's crosses ($\times$), and Greek crosses (+), respectively.
(The symbols for model~0, models~1--3, models~4--6, and models~7--9 are coloured in
black, blue, green, and red, respectively.) 
}
\label{fig3}
%\end{minipage}
\end{figure}

\subsection{Fourier transform and the trend of first zero}

By using such corrected abundances, the theoretical line profiles were simulated 
for each of the 10 models and their Fourier transform were computed, from which 
$q_{1}^{j}$ and $q_{2}^{j}$ ($j = 0, 1, 2, \cdots, 9$) were measured.
These $q_{1}^{j}$ and $q_{2}^{j}$ values along with the adopted abundance corrections 
($\Delta A^{j}$) for all 90 lines are given in ``calparms.dat'' of the supplementary material.

As demonstrative examples, the simulated profiles of Fe~{\sc i} 5133.681 ($K = -13.76$) 
and Fe~{\sc ii} 4951.584 ($K = +2.92$) lines and their Fourier transform amplitudes, 
which were calculated for models 0, 1, 3, 5, 7, and 9, are illustrated in Fig.~4, where 
the observed data are also overplotted for comparison. 
It can be seen from Fig.~4 that the behaviours of zero frequency for these two 
lines of different $K$ are just the opposite in the sense that $q_{1}$ of 
Fe~{\sc i}~5133.681/Fe~{\sc ii}~4951.584 moves towards lower/higher direction 
as the gravity-darkening effect is enhanced with an increase in $v_{\rm e}$. 

From now on, our discussion focuses only on the first zero frequency ($q_{1}$),
which is less affected by measurement errors or noises in comparison to $q_{2}$.
In order to elucidate the trend of $q_{1}$ as a function of $K$ and $v_{\rm e}$,
the $q_{1}$ values are plotted against $K$ in Fig.~5a--5f (each corresponding to 
models 0, 1, 3, 5, 7, and 9, respectively).
Besides, the linear regression lines (determined from the $q_{1}$ vs. $K$ plots 
for Fe~{\sc i} and Fe~{\sc ii} lines separately) are also shown in each panel, 
and these regression lines for all models are depicted together in Fig.~5g.
An inspection of Fig.~5 reveals the following characteristics.
\begin{itemize} 
\item
$q_{1}$ generally increases with an increase in $K$, which was already mentioned 
in Sect.~2.2 in reference to Fig.~2b. The $q_{1}$ values for Fe~{\sc i} lines are 
generally smaller than those of Fe~{\sc ii} lines because of the difference in $K$.
\item
The slope of the $q_{1}$ vs. $K$ plots is a systematic function of $v_{e}$;
i.e., it becomes progressively steeper with an increase in $v_{\rm e}$ (Fig.~5g).
This is a useful property for estimating $v_{\rm e}$ from the observed
$q_{1}$--$K$ relation. 
\item
The sensitivity of $q_{1}$ to a change in $v_{\rm e}$ also depends upon $K$ 
(cf. Fig.~5h). While $\langle {\rm d}q_{1}/{\rm d}v_{\rm e}\rangle \la 0$ holds 
for most lines of $K \la 0$ (all Fe~{\sc i} lines and many Fe~{\sc ii} lines), 
a group of high-excitation Fe~{\sc ii} lines ($\chi_{\rm low} \sim 10$~eV;
such as Fe~{\sc ii}~4951.584 in Fig.~4) with positive $K$ indicate
$\langle {\rm d}q_{1}/{\rm d}v_{\rm e}\rangle > 0$.   
\end{itemize}

%Fig. 4
\setcounter{figure}{3}
\begin{figure*}
\begin{minipage}{120mm}
\includegraphics[width=12.0cm]{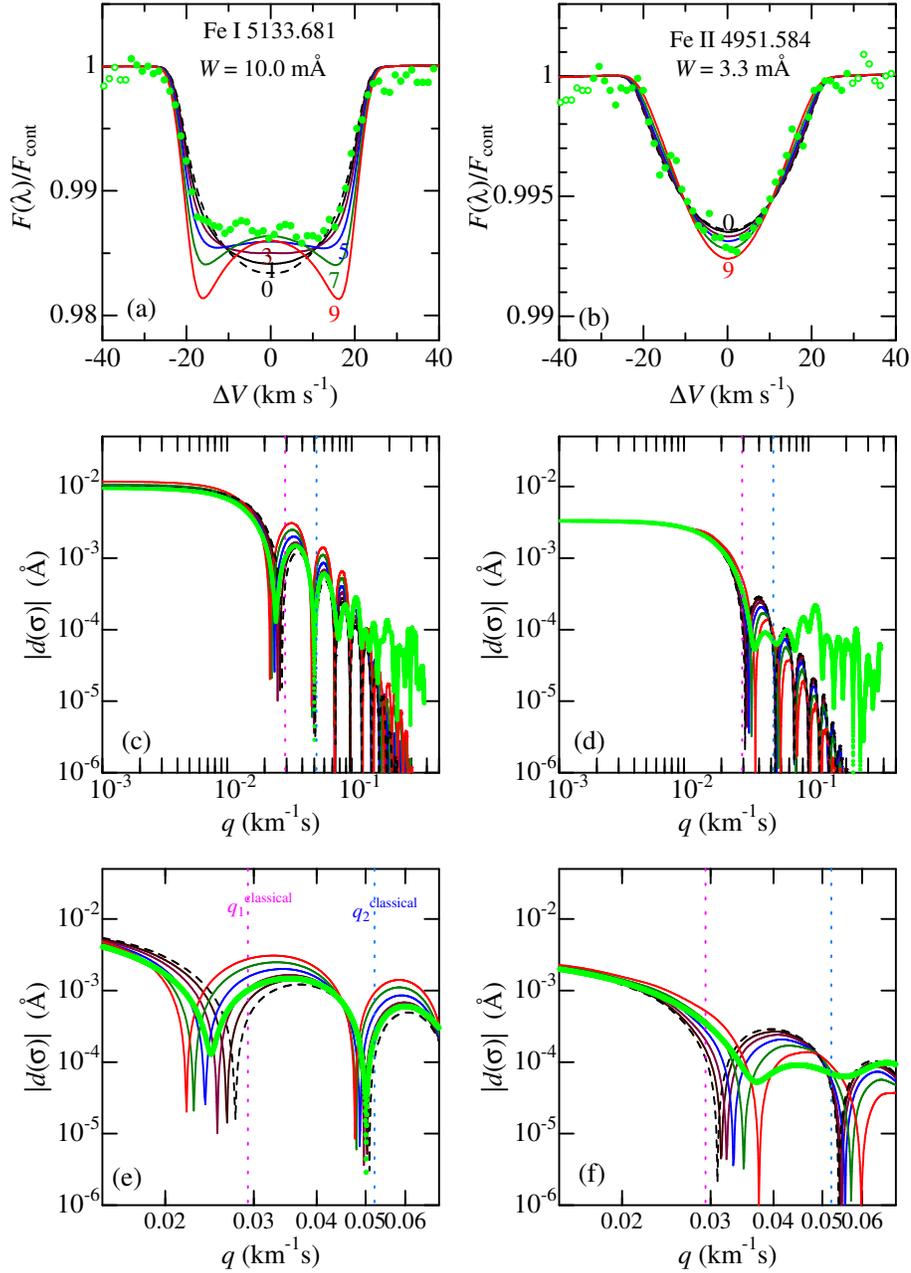}
\caption{
Theoretical line profiles (top panels) and their Fourier transform amplitudes 
(middle/bottom panels for wide/zoomed view) of Fe~{\sc i} 5133.681 (left-hand side) and 
Fe~{\sc ii} 4951.584 (right-hand side) calculated for models 0 (black dashed line), 
1 (black solid line), 3 (purple solid line), 5 (blue solid line), 7 (green solid line), 
and 9 (red solid line), while the observed data are also overplotted by light-green symbols.
In the middle/bottom panels, the positions of $q_{1}^{\rm classical}$ and $q_{2}^{\rm classical}$ 
corresponding to the classical rotational broadening function (cf. Eqs.~3, 4, and 6 in Paper~II)
are indicated by vertical dotted lines for comparison. 
}
\label{fig4}
\end{minipage}
\end{figure*}

%Fig.5
\setcounter{figure}{4}
\begin{figure*}
\begin{minipage}{140mm}
\begin{center}
\includegraphics[width=14.0cm]{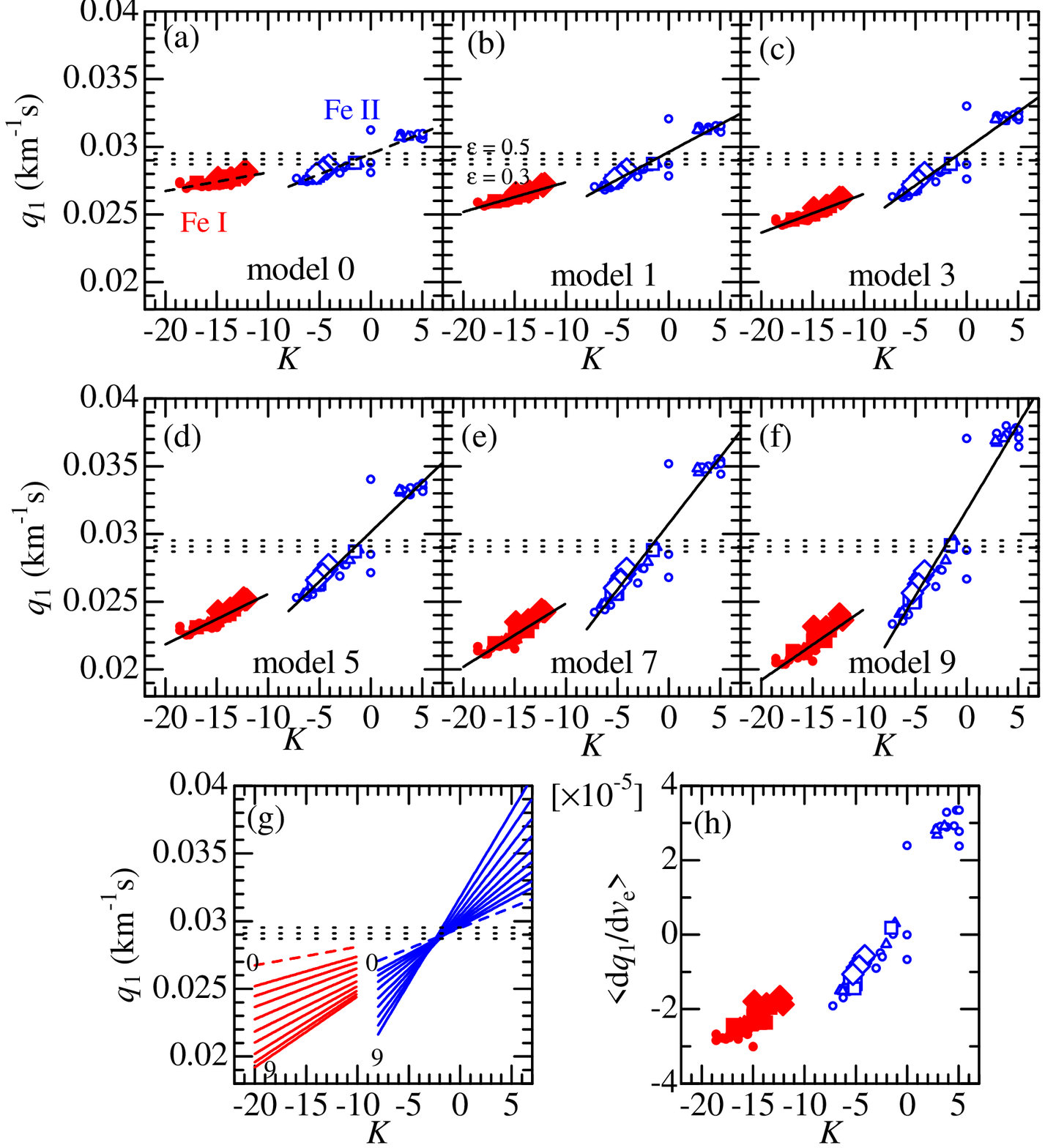}
\caption{ 
Panels (a)--(f) show the simulated relationships between $q_{1}$ (1st-zero frequency) 
and $K$ ($T$-sensitivity parameter) for the 49 Fe~{\sc i} (red filled symbols) 
and 41 Fe~{\sc ii} lines (blue open symbols) calculated for models 0, 1, 3, 5, 7, 
and 9, respectively. The size and shape of the symbols denote the difference 
in line strengths; circles $\cdots$ $W^{\rm obs} <$~5~m\AA, 
triangles $\cdots$ 5~m\AA~$\le W^{\rm obs} <$~10~m\AA,
squares $\cdots$ 10~m\AA~$\le W^{\rm obs} <$~20~m\AA, and
diamonds $\cdots$ 20~m\AA~$\le W^{\rm obs} <$~40~m\AA.
The linear regression lines derived from these $q_{1}$ vs. $K$ plots 
(separately for Fe~{\sc i} and Fe~{\sc ii}) are also overplotted by solid lines in each 
of the panels (a)--(f), and those for all 10 models are put together in panel (g).
The three horizontal dotted lines represent the classical $q_{1}$ values (corresponding 
to $v_{\rm e}\sin i$ = 22~km~s$^{-1}$) for the limb-darkening coefficient 
($\epsilon$) of 0.3, 0.4, and 0.5 (cf. Eq.~3 in Paper~II). 
In panel (h) are plotted the mean gradients $\langle {\rm d}q_{1}/{\rm d}v_{\rm e}\rangle$ 
(in unit of km$^{-2}$s$^{2}$; averaged over $v_{\rm e}$ between 100 and 300~km~s$^{-1}$) 
against $K$, which were computed from the coefficients of quadrature fit 
($q_{1} = A + B v_{\rm e} + C v_{\rm e}^{2}$) as $B+2C \times (100+300)/2$ (i.e., 
${\rm d}q_{1}/{\rm d}v_{\rm e}$ at the mid-$v_{\rm e}$).
}
\label{fig5}
\end{center}
\end{minipage}
\end{figure*}

%Sect. 4
\section{Result and discussion}

\subsection{Rotational velocity of Vega}

Now that the observational data of zero frequencies ($q_{1}^{\rm obs}$) as well as 
the corresponding theoretically calculated values ($q_{1}^{{\rm cal},j}$ for 
$j = 0, 1, \cdots, 9$) to be compared are all set for 90 lines, we can address 
the main task of investigating Vega's rotational velocity, while following 
the same procedure as adopted in Paper~II (cf. Sect.~4.3 therein).

The observed $q_{1}^{\rm obs}$ values  are plotted against $K$ in Fig.~6. 
As seen from this figure, these $q_{1}^{\rm obs}$ data show an increasing tendency 
with $K$ and those for Fe~{\sc i} and Fe~{\sc ii} lines are distributed in separate 
two groups, which is quite similar to the theoretical predictions mentioned in Sect.~3.3 
(cf. Figs.~5a--5f). Therefore, there is a good hope of successfully establishing 
$v_{\rm e}$ by comparing $q_{1}^{\rm obs}$ and $q_{1}^{{\rm cal}}$ for many lines altogether.

%Fig.6
\setcounter{figure}{5}
\begin{figure}
\begin{minipage}{80mm}
\begin{center}
\includegraphics[width=8.0cm]{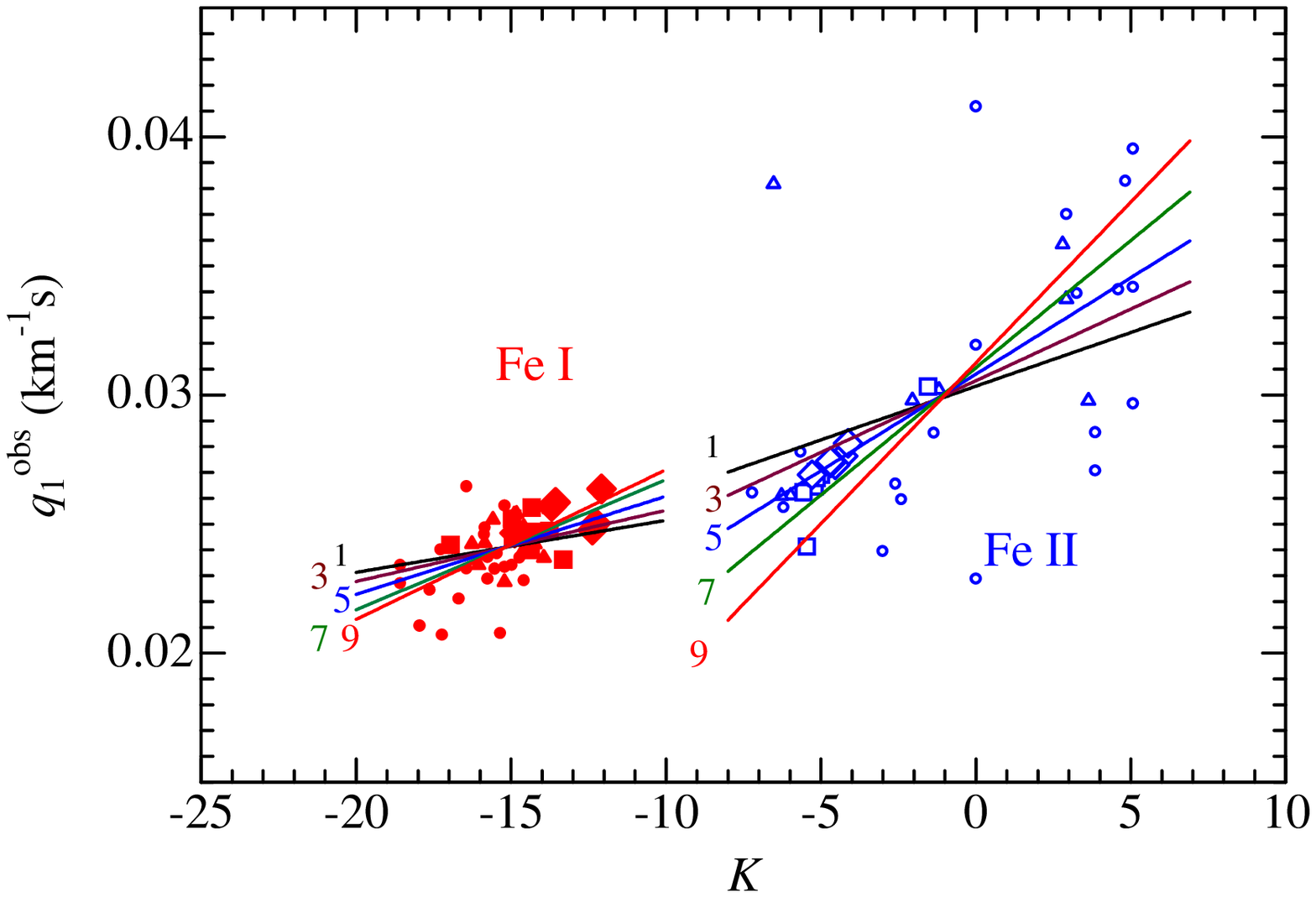}
\caption{
Observed $q_{1}$ values of Fe~{\sc i} and Fe~{\sc ii} lines plotted against $K$,
where the meanings of the symbols are the same as in Fig.~5.
The averaged trends (gradients) of theoretical $q_{1}$ vs. $K$ relations calculated 
for models 1, 3, 5, 7, and 9 (determined by linear-regression analysis; cf. Fig.~5) 
are also depicted by solid lines, which were multiplied by a scaling factor of 
$22/x^{*}$ in order to adjust the difference between the actual 
$v_{\rm e}\sin i (\equiv x)$ and the assumed value (22~km~s$^{-1}$) in the profile
calculation (see Table~4 for the $v_{\rm e}$-dependent values of $x^{*}$).
}
\label{fig6}
\end{center}
\end{minipage}
\end{figure}

Since the actual value of $v_{\rm e}\sin i$ (hereinafter denoted as $x$ for simplicity) 
is likely to be slightly different from 22~km~s$^{-1}$ assumed for calculating the 
modelled profiles, $q_{1}^{{\rm th},j}$ should be multiplied by a scaling factor ($22/x$) 
to allow for this possible difference.
The standard deviation defined as 
\begin{equation}
\sigma(x^{i},v_{\rm e}^{j}) \equiv 
\sqrt{ 
\frac{ \sum_{n=1}^{N} [q_{1}^{\rm obs}(n) - q_{1}^{{\rm cal},j}(n)(22/x^{i})]^{2}}{N}.
}
\end{equation}
was computed for each combination of ($x^{i}$, $v_{\rm e}^{j}$),
where $x^{i} = 15.0 + 0.2i$ ($i$ = 0, 1, $\cdots$, 75) and $v_{\rm e}^{j} = 100 + 25(j-1)$ 
($j$ = 1, 2, $\cdots$, 9). Here, $n$ is the index of each line and $N$ is the total 
number of the lines used. As in Paper~II, Fe~{\sc i} lines ($N = 49$) and Fe~{\sc ii} 
lines ($N = 41$) are treated separately.
The best ($x$, $v_{\rm e}$) solution may be found by searching for the location of 
$\sigma$ minimum. 

The behaviours of the resulting $\sigma$ (3D surface and contour plots) are displayed 
in Fig.~7 (left and right panels are for Fe~{\sc i} and Fe~{\sc ii}, respectively).
The trace line connecting ($x^{*}$, $v_{\rm e}$) is also overplotted by the dashed 
line, where $x^{*}$ corresponds to the minimum of $\sigma$ trough for each $v_{\rm e}$ 
(in Table~4 are given the actual data of $x^{*}$ and the corresponding $\sigma^{*}$). 
Besides, the run of $\sigma$ with $v_{\rm e}$ across the tracing is depicted 
in Fig.~8a, and the tracings for both species are drawn together in Fig.~8b. 

%Fig.7
\setcounter{figure}{6}
\begin{figure*}
\begin{minipage}{120mm}
\begin{center}
\includegraphics[width=12.0cm]{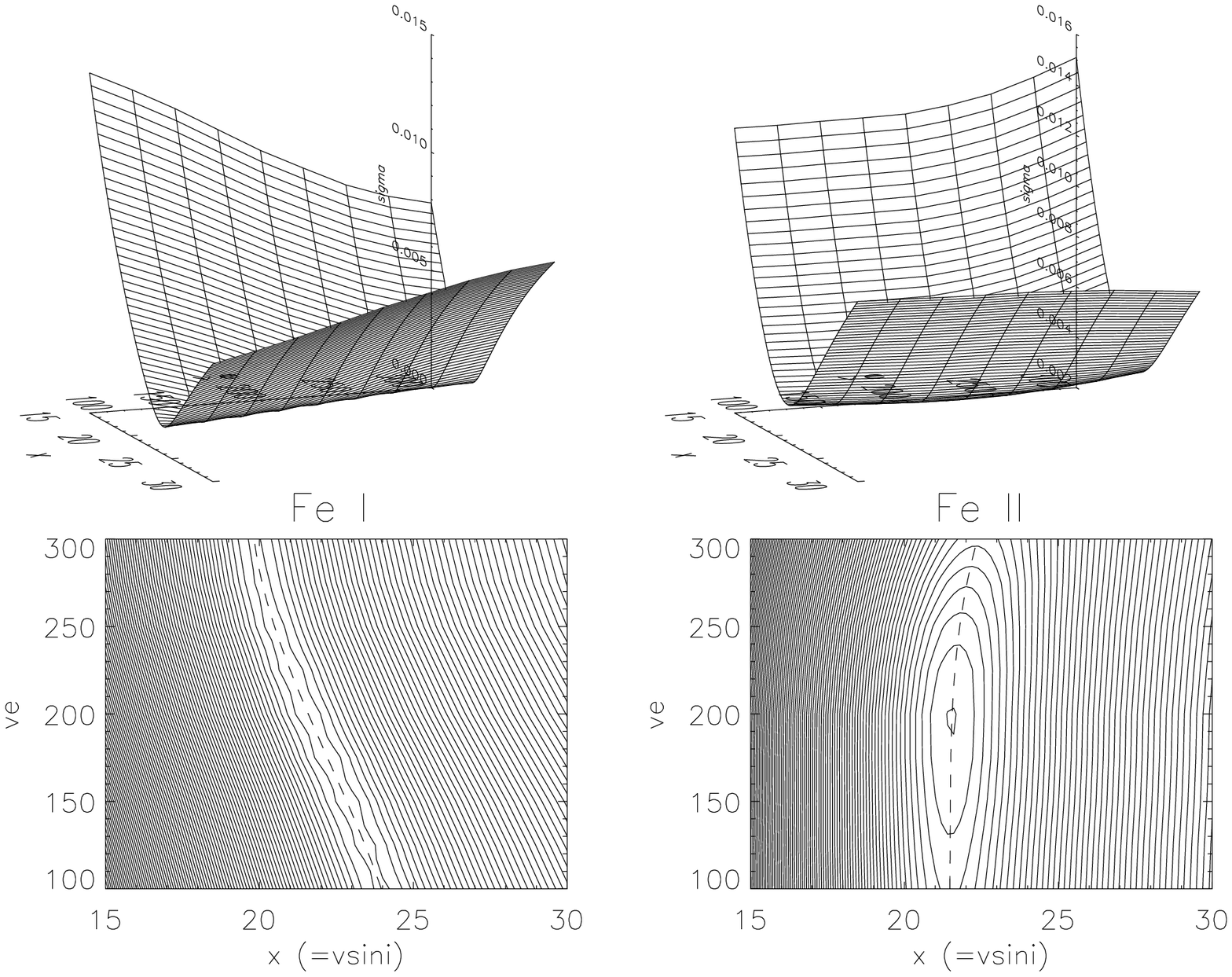}
\caption{
Graphical display of the behaviour of $\sigma$, which is the standard deviation 
between the simulated $q_{1}^{\rm cal}(x,v_{\rm e})$ (where $x\equiv v_{\rm e}\sin i$) and 
the observed $q_{1}^{\rm obs}$ for each of the Fe lines, where the results for 
Fe~{\sc i} and Fe~{\sc ii} lines are separately displayed in the left and 
right, respectively. Each set consists of the 3D representation of the $\sigma(x,v_{\rm e})$ 
surface (upper panel) and the contours of $\sigma$ on the $x$--$v_{\rm e}$ plane (lower panel).
The trace of trough bottom (connection of $x^{*}$ values at the minimum $\sigma$ for each given 
$v_{\rm e}$; cf. Table~4) is indicated by the dashed line in the contour plot.
}
\label{fig7}
\end{center}
\end{minipage}
\end{figure*}

%Fig.8
\setcounter{figure}{7}
\begin{figure}
\begin{minipage}{80mm}
\begin{center}
\includegraphics[width=8.0cm]{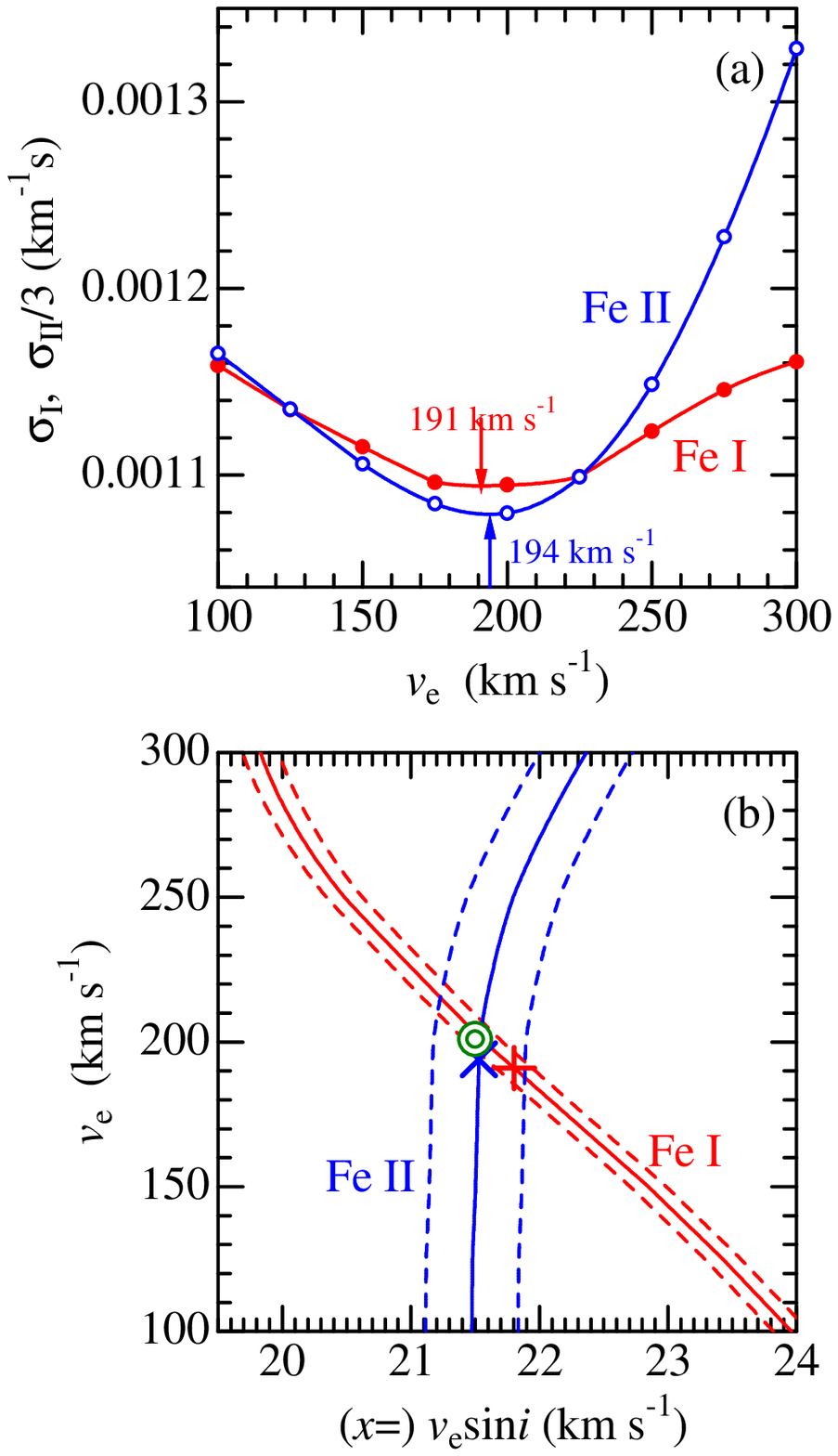}
\caption{
(a) $\sigma$ vs. $v_{\rm e}$ relation along the trough bottom 
for Fe~{\sc i} (filled symbols) and Fe~{\sc ii} (open symbols); in each case, 
the position of minimum $\sigma$ (evaluated by interpolation) is indicated by an arrow.
Note that $\sigma_{\rm II}$ ($\sigma$ for Fe~{\sc ii}) is reduced by a factor 
of 1/3 in this figure.
(b) The traces of trough bottoms for Fe~{\sc i} and Fe~{\sc ii} (dashed lines in 
the contour panels of Fig.~7) are plotted together in the $v_{\rm e}$ vs. $x$ plane 
by solid lines (the intersection is shown by the double circle), 
while the dashed lines indicate the error bars involved in $x$ 
($\pm 0.14$~km~s$^{-1}$ for Fe~{\sc i} and $\pm 0.36$~km~s$^{-1}$ for Fe~{\sc ii}).
The minimum positions of $\sigma_{\rm I}$ and $\sigma_{\rm II}$ are also indicated
by Greek cross (+) and St. Andrew's cross ($\times$), respectively.
}
\label{fig8}
\end{center}
\end{minipage}
\end{figure}

An inspection of Fig.~8 yielded satisfactory results, because three kinds 
of ($x$, $v_{\rm e}$) solutions turned out consistent with each other: (191, 21.8)
from the minimum of $\sigma_{\rm I}$ (Fig.~8a), (194, 21.5) from the minimum of 
$\sigma_{\rm II}$ (Fig.~8a), and (201, 21.5) from the intersection of two trace lines (Fig.~8b). 

The uncertainties involved in $x^{*}$ were estimated as $\sim 0.14$~km~s$^{-1}$ (Fe~{\sc i})
and  $\sim 0.36$~km~s$^{-1}$ (Fe~{\sc ii}),\footnote{
This estimation is based on the relation $\delta x/x \sim \delta q_{1}/q_{1}$,
where $\delta q_{1} \sim \sigma /\sqrt{N}$. See Sect.~4.3 in Paper~II for more details.} 
which are indicated by dashed lines in Fig.~8b. From this figure, errors in $v_{\rm e}$ 
and $x$ were roughly evaluated (from the size of the parallelogram area embraced by 
4 dashed lines around the intersection) as $\sim \pm 15$~km~s$^{-1}$
and $\sim \pm 0.3$~km~s$^{-1}$, respectively.

Consequently, by averaging these three solutions, Vega's equatorial and projected 
rotational velocities were concluded as $v_{\rm e} = 195 (\pm 15)$~km~s$^{-1}$ and 
$v_{\rm e}\sin i = 21.6 (\pm 0.3)$~km~s$^{-1}$, which further result in 
$i = 6.4^{\circ} (\pm 0.5^{\circ})$.
Among the 10 models adopted in this study (cf. Table~3), model~5 ($v_{\rm e} = 200$~km~s$^{-1}$)
is the most preferable model; this can be actually confirmed in Fig.~6, where the linear-regression 
lines defined in Fig.~5b--5f are overplotted (after the $v_{\rm e}$-dependent difference 
between $x^{*}$ and 22 has been corrected). 

%Table 4
\setcounter{table}{3}
\begin{table}
%\begin{minipage}{160mm}
\small
%\scriptsize
\caption{Behaviours of $\sigma$ trough for Fe~{\sc i} and Fe~{\sc ii} lines.}
\begin{center}
\begin{tabular}{c@{ }c@{ }cccc}\hline
Model & $v_{\rm e}$ & $x^{*}_{\rm I}$ & $x^{*}_{\rm II}$ & $\sigma^{*}_{\rm I}$ &  $\sigma^{*}_{\rm II}$ \\
number & (km~s$^{-1}$) & (km~s$^{-1}$) & (km~s$^{-1}$) & (km$^{-1}$s) & (km$^{-1}$s) \\ 
\hline
1 & 100 & 23.9601 & 21.4741 & 0.0011587 & 0.0034953 \\
2 & 125 & 23.4263 & 21.4789 & 0.0011351 & 0.0034061 \\
3 & 150 & 22.8475 & 21.5036 & 0.0011152 & 0.0033180 \\
4 & 175 & 22.2098 & 21.5175 & 0.0010961 & 0.0032541 \\
5 & 200 & 21.5720 & 21.5339 & 0.0010948 & 0.0032388 \\
6 & 225 & 21.0187 & 21.6385 & 0.0010991 & 0.0032970 \\
7 & 250 & 20.4830 & 21.7954 & 0.0011235 & 0.0034459 \\
8 & 275 & 20.0921 & 22.0538 & 0.0011457 & 0.0036829 \\
9 & 300 & 19.8300 & 22.3655 & 0.0011607 & 0.0039852 \\
\hline
\end{tabular}
\end{center}
These data show the characteristics of the trough in the $\sigma(x,v_{\rm e})$ surface 
($x \equiv v_{\rm e}\sin i$) defined by Eq.~2 for each group of Fe~{\sc i} and Fe~{\sc ii} lines.
$x^{*}$ is the $x$ value at the minimum $\sigma(x,v_{\rm e})$ for each given $v_{\rm e}$,
and $\sigma^{*}$ is the corresponding $\sigma(x^{*},v_{\rm e})$. The trace of $x^{*}$ as a
function of $v_{\rm e}$ is shown by the dashed line in the contour plot of Fig.~7.
%\end{minipage}
\end{table}

\subsection{Comparison with previous results}

As mentioned in Sect.~1, although the considerably large differences of Vega's $v_{\rm e}$
amounting to $\ga 100$~km~s$^{-1}$ seen in the literature of early time were reduced 
in the more recent results (most of them were published within several years around 2010),
they are still diversified ranging from $\sim 170$ to $\sim 230$~km~s$^{-1}$.
Interestingly, the $v_{\rm e}$ value ($\sim 200$~km~s$^{-1}$) derived in this study 
is almost in-between this dispersion.
It may be worth briefly reviewing these literature $v_{\rm e}$ values (published 
since Paper~I; cf. Table~1) in comparison with the consequence of this investigation.  
\begin{itemize}
\item
Line profile method:\\
Paper~I's result (175~km~s$^{-1}$) based on the conventional profile fitting has been 
revised upward by $\sim +20$~km~s$^{-1}$ in this reinvestigation by applying the Fourier 
transform method to line profiles.
While Yoon et al.'s (2010) 236~km~s$^{-1}$ is somewhat too large, Hill, Gulliver \& 
Adelman's (2010) 211~km~s$^{-1}$ is in tolerable agreement as compared with the preset result.    
\item
Interferometry method:\\
Monnier et al.'s (2012) conclusion of $v_{\rm e} = 197$~km~$^{-1}$ (derived from 
$v_{\rm e}\sin i$ and $i$ given in their Table~2 as Model~3) based on optical 
interferometry is in good agreement with this study. Actually, 
Fig.~2 of Monnier et al. (2012) shows that their Model~3 matches well with
model~5 ($v_{\rm e} = 200$~km~s$^{-1}$) of Paper~I.
\item
Magnetic modulation method:\\
Vega's rotational period ($P$) was directly determined by way of detecting the magnetic 
modulation based on time-sequence data of spectropolarimetric observations:
0.732~d (Petit et al. 2010), 0.678~d (Alina et al. 2012), 0.623~d (Butkovskaya 2014),
and 0.678~d (B\"{o}hm et al. 2015).
Among these four, it is the $P$ value of 0.678~d derived by both Alina et al. and 
B\"{o}hm et al. that is most consistent with the $v_{\rm e}$ result (195~km~s$^{-1}$) 
of this investigation, which corresponds to $P = 2\pi R_{\rm e}/v_{\rm e} = 0.685$~d 
(where $R_{\rm e} = 2.784$~R$_{\odot}$
for model~5 is adopted).
\end{itemize}

\subsection{Advantage of Fourier analysis}

Finally, some comments may be in order regarding the superiority of exploiting
the zero frequency ($q_{1}$) measured from the Fourier transform of line profiles  
in comparison with the ordinary profile fitting approach in the wavelength domain.

The distinct merit of using $q_{1}$ is that it can discern very subtle differences
in the profile shape. Fig.~4 provides a good demonstrative example.
While the profile of Fe~{\sc i} 5133.681 undergoes a comparatively easy-to-detect change 
with an increase in $v_{\rm e}$ (Fig.~4a), that of Fe~{\sc ii} 4951.584 is apparently
inert (Fig.~4b), which means that getting information on $v_{\rm e}$ 
from the profile of the latter is more difficult (this is the reason why Fe~{\sc ii} 
lines could not be used for determining $v_{\rm e}$ in Paper~I). However, the situation 
is different in the Fourier space, where the shift of $q_{1}$ (reflecting the change of 
line profile) is sufficiently detectable with almost the same order of magnitude for 
both cases (cf. Figs.~4e and 4f). Accordingly, Fe~{\sc i} as well as Fe~{\sc ii} lines 
are equally usable for $v_{\rm e}$ determination if $q_{1}$ is invoked, as done 
in this study.  

Besides, $q_{1}$ is precisely measurable and easy to handle as a single parameter, 
which is a definite advantage from a practical point of view. Actually, $q_{1}$ data 
of many lines can be so combined as to improve the precision of $v_{\rm e}$ (while 
statistically estimating its error) as done in this paper. Such a treatment would be 
difficult in the conventional approach of fitting the observed and theoretical profiles.

\subsection{Line profile classification using $q_{1}$ and $K$}

Another distinct merit of $q_{1}$ is that it provides us with a prospect
for quantitative classification of spectral line shapes founded on a 
physically clear basis. Since the discovery around $\sim$~1990 that 
a number of spectral lines in Vega (e.g., weak lines of neutral species) 
show unusual profiles of square form, there has been a tendency to pay 
attention to this specific line group (e.g., compilation of flat-bottomed 
lines in Vega by Monier et al. 2017). However, the actual situation of 
Vega's spectral lines in general is not so simple as to be dichotomised 
into two categories of normal and peculiar profiles; as a matter of fact, 
the individual profiles of most lines should more or less have anomalies 
of different degree. Unfortunately, detection of such details has been 
hardly possible so far, because the judgement of profile peculiarity was 
done by simple eye-inspection due to the lack of effective scheme for 
describing/measuring the delicate characteristics of line profiles. 

The first zero frequency ($q_{1}$) is just what is needed in this context, 
which is not only sensitive to a slight difference of line shape but 
also easily measurable in the Fourier space. Moreover, thanks to its close 
relationship with $K$, the behaviour of $q_{1}$ (representing the line shape 
characteristics) can be reasonably explained in terms of the underlying 
physical mechanism. We now have a unified understanding as to why different 
spectral lines exhibit diversified profiles in Vega, as summarised below.
\begin{itemize}
\item
It is the parameter $K$ (temperature sensitivity) that essentially determines
the observed line shape. The contribution to the important shoulder part of 
the profile away from the line centre ($|\Delta \lambda| \la \lambda v_{\rm e}\sin i / c$) 
is mainly made by the light coming from near to the 
gravity-darkened limb of lowered $T$. Accordingly, lines of $K < 0$, $K \sim 0$,
and $K > 0$ show boxy ({\bf U}-shaped), normally round (like classical rotational
broadening), and rather peaked ({\bf V}-shaped) profiles, each of which result in 
appreciably different $q_{1}$ values. For example, in Fig.~4, these three groups 
correspond to those of lower $q_{1}$ ($\sim 0.025$~km$^{-1}$s), medium 
$q_{1}$ ($\sim 0.03$~km$^{-1}$s), and higher $q_{1}$ ($\sim 0.035$~km$^{-1}$s), 
respectively. 
\item
The peculiarity degree of the line shape (i.e., departure from the classical 
rotationally-broadened profile) is described by $K$, because 
$(q_{1} - q_{1}^{\rm classical}) \propto K$ 
($q_{1}^{\rm classical} \simeq 0.03$~km$^{-1}$s) and the gradient ($>0$) 
of this relation progressively increases with $v_{\rm e}$, 
as manifested in Fig.~5. As such, the profile of any line in Vega can be 
reasonably predicted if $K$ and $v_{\rm e}$ are specified. 
\item
As explained in Appendix~A of Paper~II, the value of $K$ for each spectral line 
depends upon $\chi_{\rm low}$ (lower excitation potential) and $W$ (equivalent 
width). It is important to note that the line strength affects $K$ in the sense 
that $|K|$ tends to decrease with an increase in $W$ (i.e., as the line becomes
more saturated), which means that chemical abundances are implicitly involved. 
In the present case of A-type stars, $K$ values for Fe~{\sc i} lines are determined
mainly by $W$ while those for Fe~{\sc ii} lines are primarily by $\chi_{\rm low}$ 
(cf. Fig.~A1 in Paper~II), which are also indicated from Fig.~2c and Fig.~2d.
\item
These behaviours of $K$ in terms of the line parameters reasonably explain why 
different spectral lines of Vega reveal various characteristic shapes. For example: 
(1) Flat-bottom profiles (manifestation of $K<0$) are seen in Fe~{\sc i} lines 
but not in Fe~{\sc ii} lines, because of the distinct difference in $K$ between these 
two line groups; i.e., $-20 \la$~$K$(Fe~{\sc i})~$\la -10$ and $-5 \la K$(Fe~{\sc ii})~$\la +5$. 
(2) The reason why typical flat-bottomed shape is observed mainly in weak 
Fe~{\sc i} lines (e.g., 4707.272, 4903.308 with $W$ of several m\AA) but not 
clearly in moderate-strength Fe~{\sc i} lines (e.g., 4202.028, 4920.502 with $W$ 
of a few tens m\AA) is that the (negative) $K$ values of the former group is 
generally lower than those of the latter owing to the dependence upon $W$. 
(3) Regarding Fe~{\sc ii} lines, some lines have clearly peaked {\bf V}-shape 
(e,g., Fe~{\sc ii}~5004.195 with $\chi_{\rm low}$ = 10.272~eV and $K = +2.92$) 
while others exhibit rather rounded profile (e.g., Fe~{\sc ii}~4993.358 with 
$\chi_{\rm low}$ = 2.807~eV and $K = -6.27$), which is naturally attributed to the 
apparent distinction of $K$ (the sign is inversed) due to the large difference in 
$\chi_{\rm low}$. 
\end{itemize}

%Sect. 5
\section{Summary and conclusion}

It is known that the sharp-line star Vega ($v_{\rm e}\sin i \sim 20$~km~s$^{-1}$) 
is actually a rapid rotator seen nearly pole-on with low $i$ $(< 10^{\circ})$.
However,  its intrinsic rotational velocity is still in dispute, 
for which rather diversified values have been published. 

In the previous studies (including Paper~I by the author's group), analysis of 
spectral line profiles has been often invoked for this purpose, which contain information 
on $v_{\rm e}$ via the gravity-darkening effect, However, it is not necessarily 
easy to reliably determine $v_{\rm e}$ by direct comparison of observed and 
theoretically simulated line profiles. Besides, this approach is not methodologically 
effective because it lacks the scope for combining many lines in establishing the solution. 

Recently, the author applied in Paper~II the Fourier analysis to the profiles of 
many Fe~{\sc i} and Fe~{\sc ii} lines of Sirius~A and estimated its $v_{\rm e}$ by 
making use of the first zero ($q_{1}$) of the Fourier transform, which turned out 
successful. Therefore, the same approach was decided to adopt in this study to 
revisit the task of establishing $v_{\rm e}$ of Vega.
 
As to the observational data, the same high-dispersion spectra of Vega as adopted
in Paper~I were used. From the Fourier transforms computed from the profiles of 
selected 49 Fe~{\sc i} and 41 Fe~{\sc ii} lines, the corresponding zero frequencies 
were measured for the analysis. The $K$ values ($T$-sensitivity parameter) of
these Fe lines are in the range of  $-20 \la K \la -10$ (Fe~{\sc i} lines) and   
$-5 \la K \la +5$ (Fe~{\sc ii} lines). 

Regarding the gravity-darkened models of rotating Vega, the model grid (comprising 
10 models) arranged in Paper~I was adopted, which cover the $v_{\rm e}$ 
range of 100--300~km~s$^{-1}$ while assuming $v_{\rm e}\sin i = 22$~km~s$^{-1}$ as fixed. 
The theoretical profiles of 90 lines were simulated for each model, from which  
Fourier zero frequencies were further evaluated. 

An inspection of these $q_{1}^{\rm cal}$ values for the simulated profiles revealed 
an increasing tendency with $K$ and the slope of this trend becomes steeper towards 
larger $v_{\rm e}$, which suggests that $v_{\rm e}$ is determinable by comparing
$q_{1}^{\rm cal}(K,v_{\rm e})$ with observed $q_{1}^{\rm obs}$ for many lines of different $K$.

It turned out that $v_{\rm e}$ and $v_{\rm e}\sin i$ could be separately established 
by the requirement that the standard deviation of the residual between $q_{1}^{\rm cal}$ 
and $q_{1}^{\rm obs}$ be minimised (while taking into account the difference between
the actual $v_{\rm e}\sin i$ and 22~km~s$^{-1}$ assumed in the model profiles),
and independent analysis applied to two sets of Fe~{\sc i} and Fe~{\sc ii} 
lines yielded solutions consistent with each other. 

The final parameters of Vega's rotation were concluded to be  
$v_{\rm e}\sin i = 21.6 (\pm 0.3)$~km~s$^{-1}$,
$v_{\rm e} = 195 (\pm 15)$~km~s$^{-1}$, and $i = 6.4 (\pm 0.5)^{\circ}$.

\section*{Acknowledgements}
This research has made use of the SIMBAD database, operated by CDS, 
Strasbourg, France. 

\section*{Data availability}

The data underlying this article are available in the 
supplementary materials.

\section*{Supporting information}

Additional Supporting Information may be found in the supplementary materials.
\begin{itemize}
\item
{\bf ReadMe.txt} 
\item
{\bf obsparms.dat} 
\item
{\bf calparms.dat} 
\item
{\bf obsprofs.dat} 
\end{itemize}
Please note: Oxford University Press is not responsible for the
content or functionality of any supporting materials supplied by
the authors. Any queries (other than missing material) should be
directed to the corresponding author for the article.

\end{document}